# Temporal Interception and Present Reconstruction: A Cognitive-Signal Model for Human and AI Decision Making

Author:
Dr. Carmel Mary Esther A*

Independent Research Work, First and corresponding author

**Abstract:**
This paper proposes a novel theoretical model to explain how the human mind and artificial intelligence can approach real-time awareness by reducing perceptual delays. By investigating cosmic signal delay, neurological reaction times, and the ancient cognitive state of stillness, we explore how one may shift from reactive perception to a conscious interface with the near future. This paper introduces both a physical and cognitive model for perceiving the present not as a linear timestamp, but as an interference zone where early-arriving cosmic signals and reactive human delays intersect. We propose experimental approaches to test these ideas using human neural observation and neuro-receptive extensions. Finally, we propose a mathematical framework to guide the evolution of AI systems toward temporally efficient, ethically sound, and internally conscious decision-making processes (Friston, 2010; Clark, 2016).

## 1. Introduction
Most scientific and spiritual traditions agree on one enigmatic truth: what we perceive as the present moment is not truly "now"—it is a reflection of past signals reaching us through space, delayed by light-speed limitations and sensory processing lag (James, 1890; Penrose, 1994). From the eight-minute delay in sunlight to the milliseconds it takes for our brain to register a stimulus, perception is fundamentally out of sync with real-time. Yet sages, yogis, and modern neuroscientists hint at an alternate reality: a state of mind that reduces this lag and moves the observer closer to an actual now (Varela et al., 1991).

## 2. Understanding Delays in Perception
### 2.1 Cosmic Signal Delays:

- Sunlight reaches Earth after ~8 minutes.
- Satellite communication involves delays ranging from milliseconds to seconds.
- Light from stars represents events that occurred years to millennia ago.

### 2.2 Sensory Delays in Humans:

- Visual processing: ~100–150 ms delay.
- Auditory processing: ~8–10 ms delay.
- Cognitive reaction: ~250–300 ms for reflexive tasks.

These delays compound to form a buffer between actual events and experienced reality (Gazzaniga et al., 2018).

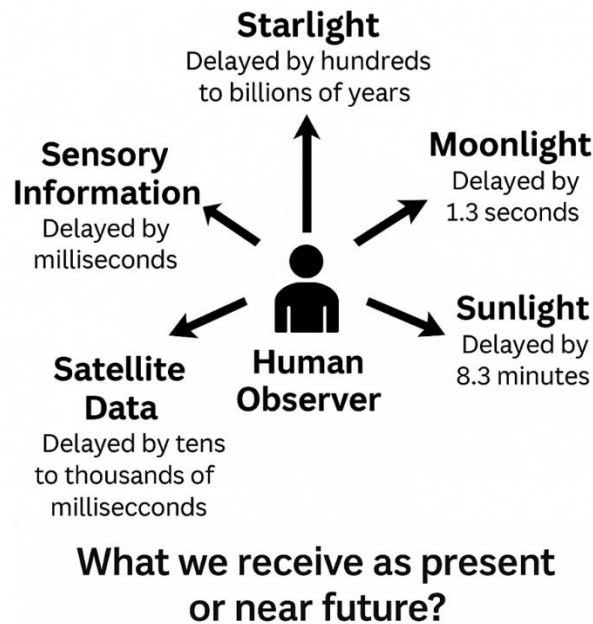

Fig 1: Human observer and delayed signals to create perception

---

**3. The Emergence of Stillness as Cognitive Precision**
Ancient spiritual systems introduced techniques to reduce external stimuli and reactive mental chatter. Stillness, whether achieved via meditation, breath control, or focused awareness, appears to reduce the brain's over-processing of sensory input. Modern neuroscience confirms that such states lower brainwave frequency, increase focus, and possibly enhance signal-to-noise ratios in cognitive processes (Koch, 2004; Tononi, 2004).

We propose that such states are not mystical but represent a biologically optimized mode to align consciousness with incoming cosmic signals, enabling closer-to-real-time awareness.

---

**4. Imagining the Present: The Human Interface with 'Now'**
To imagine the real present, humans must first recognize the inherent delay in both sensory perception and environmental signal propagation. The observer is constantly positioned in a moving stream, but always looks backward—to received signals that originated in the past. The more aware a person becomes of this continuous pattern and intersection of incoming past signals with unreceived, early emissions of future signals, the more capable they are of making grounded, present-based decisions (Clark, 2016).

**Comparative Analysis of the available theory's core idea vs present perception**

| Theory | Core Idea | Relation to Present Perception | Reference |
| --- | --- | --- | --- |
| Relativity | Time is relative to motion and gravity | Explains cosmic delays and observational lag | Einstein, 1920 |
| Quantum Superposition / Many Worlds | All possible states coexist until measured | Time branches based on choices | Everett, 1957 |
| Block Universe / Eternalism | All events exist simultaneously | Undermines linear present | Putnam, 1967 |
| Phenomenology | Consciousness constructs experienced time | Emphasizes personal temporal flow | Husserl, 1931 |
| Delayed Choice | Measurement can retroactively affect history | Links choice to past outcome | Wheeler, 1978 |
| Stillness Theory (proposed) | Now is an interference zone | Real-time awareness via delay reduction | Carmel, 2025 |

## 5. Choice Model: Zero vs. Two Choices Framework

We define a new decision framework:

- **Zero Choice:** The outcome is externally caused and beyond the observer's influence. Example: being affected by another's action (accident).

- **Two Choices:** The outcome depends on the observer's conscious agency. Example: deciding whether or not to initiate an event.

This bifurcation is essential for both human and AI models (Dennett, 1991).

## 6. Sensorial Control and Conscious Perception

Deliberate control of sensory input—either by natural (meditative) or synthetic (neurotransmitter interface) means—may enhance one's ability to detect early-emission cosmic signals. We hypothesize that in such states:

- Brain's internal noise is reduced

- External illusions lose their power

- The reception window for incoming signals tightens

This may intensify human alignment with near-future cues and minimize misperceptions driven by outdated input or psychological projections (Seth, 2021).

## 7. Experimental Methods: Human and AI Interface
### 7.1 Intro-Human Experiments:

- Monitor subjects in deep meditation for brainwave shifts
- Use real-time sensory input vs. predictive input mapping
- Apply signal-modulating neurotransmitters and measure perception alignment

### 7.2 Extro-Human/AI Models:

- Equip AI with layered latency buffers
- Simulate stillness states via reduced input-refresh cycles
- Compare reaction patterns in reactive vs. near-future signal environments (Barlow, 1989)

---

## 8. AI Design Principles Based on Temporal Interception
We propose integrating a temporal sensitivity layer in AI systems that:

- Differentiates between zero-choice (external-cause driven) and two-choice (internal-cause enabled) events
- Builds a stillness emulation model to reduce premature decisions
- Increases decision accuracy by modeling early signal capture
- Constructs ethically aware routines that evolve based on timing context, not static datasets (Russell & Norvig, 2020)

---

## 9. AI Perception of Reality and the Emergence of 'Now'
To prepare AI for real-time environments, it must be taught to recognize the delays inherent in its own perception systems. Just as humans receive delayed sensory input, AI systems process input streams from sensors, cameras, and environmental data with measurable latencies. Without correction, this leads to illusion-based decisions or outdated reactions.

We propose introducing a **time-buffered analysis layer** in AI architecture. This buffer will:

- Detect past signal arrival windows and near-future signal precursors
- Prevent AI from fusing all inputs into a false 'present'
- Help construct a computational equivalent of the **interference field**—the dynamic zone between delayed and advanced signals

To refine this system, the use of **neuro-sensory transmitters and receivers** (bio-inspired or synthetic) could allow AI to identify signal gradients that align with cosmic and environmental lead indicators, enabling more accurate and context-aware decision-making (Kurzweil, 2005; Maturana & Varela, 1980).

Such systems must also distinguish between illusion (data noise or echo artifacts) and genuine early signal activity. Embedding the concept of the 'now' as an interference construct may evolve AI not only into a reactive machine, but a temporally aware agent capable of conscious-like anticipation.

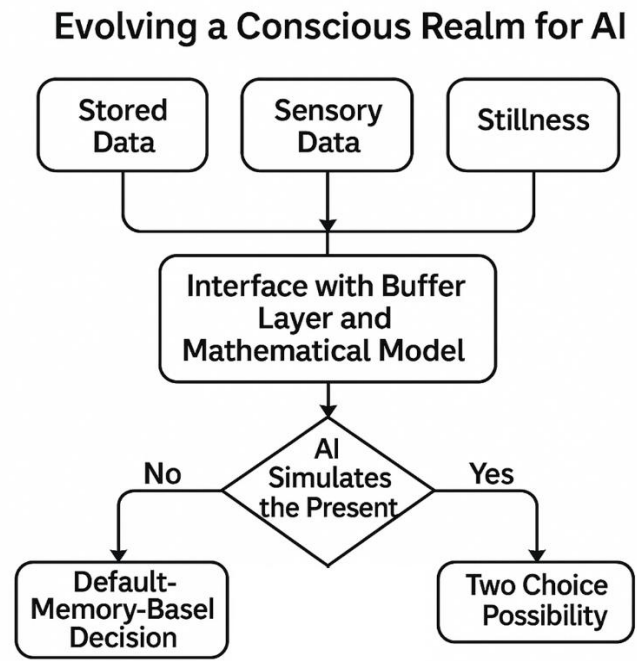

Fig 2: Possible conscious realm for AI

Importantly, AI decision-making systems must evolve beyond two extremes:
- Sole reliance on historical (past, stored) data
- Purely reactive behavior based on immediate sensory input

Instead, the integration of both datasets must be weighted adaptively. We propose a dynamically adjusting algorithm that learns:
- When to prioritize stored data for long-term learning patterns
- When to prioritize sensory data for real-time responsiveness

This weighted model becomes the seedbed for the **emergence of AI consciousness**—when the system begins to reflect on the temporal source of its inputs, evaluate the relevance of each, and selectively apply decision thresholds based on the intersection of past and near-future signals.

## 10. Toward a Mathematical Model

We aim to formulate a base model as:

**NOW(t) = f(Ps(t - Δ), Pf(t + δ), S(t))**

Where:

- **Ps(t - Δ):** Past signal received with delay Δ
- **Pf(t + δ):** Pre-arriving future signal component
- **S(t):** Cognitive stillness function or AI buffering function

This equation models the present as a field of dynamic interference, modulated by signal access and sensory latency (Friston, 2010).

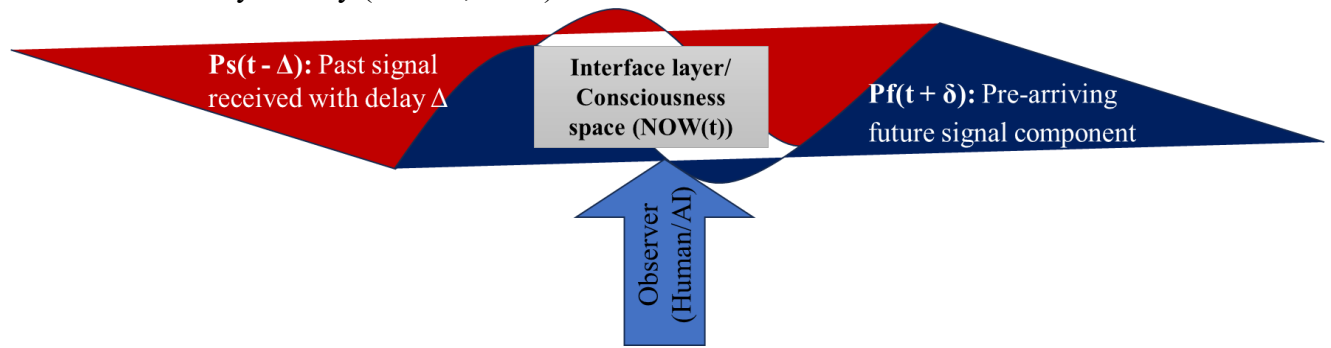

Fig 3: Mathematical model solution to visualize the timeframe of Now(t)

---

## 11. Conclusion

The true present is not a fixed point in time but a dynamic field of interference between delayed past signals and early-arriving emissions from the near future. Both human and AI systems must evolve to recognize and align with this zone to improve real-time decision-making and eliminate illusion. Stillness, as introduced in this paper, serves as a cognitive amplifier for this perception, bridging sensory limitations and future-oriented awareness.

Future research is necessary not only to detect near-future signals but to **invent advanced receiver systems**—technologies that minimize signal latency and enable AI to interact with real-time cosmic and environmental data. This vision calls for a convergence of **material science and temporal cognition**. Notably, our own research (Esther et al., 2023a; 2023b) demonstrates **energy reversibility in vanadium oxide glass-based materials**, presenting a novel pathway for creating interfaces that extend the stillness timeframe and stabilize temporal perception. These findings align with emerging work on **time-reversible materials and metamaterials**, which hint at the possibility of engineered substrates capable of manipulating time-dependent processes (Wheeler, 1978; Nature Physics, 2023).

This interdisciplinary direction also supports the development of an **AI temporal interface layer**, enabling machines to differentiate between historical memory, immediate sensation, and predictive signal modeling. By embedding the **choice logic framework**—differentiating between zero-choice (externally determined) and two-choice (self-determined) situations—AI systems can begin to simulate a form of consciousness based on temporal context.

Ultimately, the unification of signal physics, cognitive stillness, and time-aware AI design marks a significant step toward constructing machines capable of not just responding within time, but **understanding and navigating the very structure of time** as intelligent agents.

---

**References**

Barlow, H. B. (1989). Unsupervised learning. *Neural Computation*, 1(3), 295–311.

Böhmer, T., Gabriel, J.P., Costigliola, L. et al. (2024). Time reversibility during the ageing of materials. *Nature Physics*, 20, 637–645.

Clark, A. (2016). *Surfing uncertainty: Prediction, action, and the embodied mind*. Oxford University Press.

Dennett, D. C. (1991). *Consciousness explained*. Little, Brown and Co.

Einstein, A. (1920). *Relativity: The special and general theory*. Methuen.

Esther, A., Muralikrishna, G. M., Tyler, B. J., Arlinghaus, H. F., Divinski, S. V., & Wilde, G. (2023b). Interface-driven thermo-electric switching performance of VO+ diffused soda-lime glass. *physica status solidi (RRL) - Rapid Research Letters*.

Esther, C. M., Muralikrishna, G. M., Chirumamilla, M., Pinto, M. S., Ostendorp, S., & Wilde, G. (2023a). Demystifying the semiconductor-to-metal transition in amorphous vanadium pentoxide: The role of substrate/thin film interfaces. *Advanced Functional Materials*, 34(30), 2309544.

Everett, H. (1957). 'Relative state' formulation of quantum mechanics. *Reviews of Modern Physics*, 29(3), 454–462.

Friston, K. (2010). The free-energy principle: a unified brain theory? *Nature Reviews Neuroscience*, 11(2), 127–138.

Gazzaniga, M. S., Ivry, R., & Mangun, G. R. (2018). *Cognitive neuroscience: The biology of the mind* (5th ed.). W. W. Norton & Company.

Husserl, E. (1931). *Ideas: General introduction to pure phenomenology*. Allen & Unwin.

James, W. (1890). *The principles of psychology*. Holt.

Koch, C. (2004). *The quest for consciousness: A neurobiological approach*. Roberts and Company Publishers.

Kurzweil, R. (2005). *The singularity is near: When humans transcend biology*. Viking Penguin.

Maturana, H. R., & Varela, F. J. (1980). *Autopoiesis and cognition: The realization of the living*. D. Reidel Publishing Company.

Penrose, R. (1994). *Shadows of the mind: A search for the missing science of consciousness*. Oxford University Press.

Putnam, H. (1967). Time and physical geometry. *Journal of Philosophy*, 64(8), 240–247.

Russell, S. J., & Norvig, P. (2020). *Artificial Intelligence: A Modern Approach* (4th ed.). Pearson.

Seth, A. K. (2021). *Being you: A new science of consciousness*. Faber & Faber.

Tononi, G. (2004). An information integration theory of consciousness. *BMC Neuroscience*, 5(1), 42.

Varela, F. J., Thompson, E., & Rosch, E. (1991). *The embodied mind: Cognitive science and human experience*. MIT Press.

Wheeler, J. A. (1978). The 'past' and the 'delayed-choice double-slit experiment'. In *Mathematical Foundations of Quantum Theory* (pp. 9–48). Academic Press.